\documentclass[epj]{webofc}
\usepackage[varg]{txfonts}   
\woctitle{RICAP-14 The Roma International Conference on Astroparticle Physics}
\begin{document}
\title{ANTARES constraints on a Galactic component of the IceCube cosmic neutrino flux}
%
%

\author{Maurizio Spurio\inst{1,2}\fnsep\thanks{\email{maurizio.spurio@bo.infn.it}} 
}

\institute{Dipartimento di Fisica e Astronomia dell'Universit\`a di Bologna  - Viale Berti Pichat 6/2, 40127 Bologna (Italy)
\and
Istituto Nazionale di Fisica Nucleare - Sezione di Bologna - Viale Berti Pichat 6/2, 40127 Bologna (Italy)
}

\abstract{

The IceCube evidence for cosmic neutrinos has inspired a large number of hypothesis on their origin, mainly due to the poor precision on the measurement of the direction of showering events.
A North/South asymmetry in the present data set suggests the presence of a possible Galactic component. This could be originated either by single point-like sources or from an extended Galactic region. 
Expected fluxes derived from these hypotheses are presented. Some values have been constrained from the present available upper limits from the ANTARES neutrino telescope.
}
\maketitle

\section{The IceCube cosmic neutrinos and a possible Galactic component}

The recent IceCube (IC) evidence for extraterrestrial high-energy neutrinos \cite{ic1,ic2} opened new windows in the field of astroparticle physics \cite{libro}.  
With the present statistics, the High Energy Starting Events (HESE) flux observed by IC is compatible with flavor ratios $\nu_e:\nu_\mu:\nu_\tau= 1:1:1$, as expected from charged meson decays in cosmic ray (CR) accelerators and neutrino oscillation on their way to the Earth.
The non-observation of events beyond 2 PeV suggests a neutrino flux with a power law $\Phi(E) \propto E^{-\Gamma}$ with hard spectral index, e.g.  $\Gamma \simeq 2.0$, and an exponential cutoff, or an unbroken power law with a softer spectrum, e.g. $\Gamma \simeq 2.3$. 

The majority of HESE are downgoing; as the IC detector is at the South Pole, this corresponds to a larger flux from the Southern sky, where most of the Galactic plane is present. 
Table \ref{tab:ICud} (columns from 2 to 5) reports for HESE with deposited energy $E_{dep}>$ 60 TeV \cite{ic2}: the number of events; the estimated background; the number of cosmic neutrinos (i.e. the $signal$); and the number of expected cosmic $\nu$s assuming the best-fit hypothesis. Values are given separately for the North/South sky regions.

Recently, IC presented a new search for neutrinos interacting in the instrumented volume and with energy between 1 TeV and 1 PeV, using 641 days of livetime \cite{arxiv1410}. 
Table \ref{tab:ICud} (columns from 6 to 10) reports, for the events in this new sample having $ E_{dep}>25$ TeV, the same quantities defined above for the HESE.
No hypothesis test on HESE as reported in \cite{ic2} yielded at present statistically significant evidence of clustering or correlations, in particular from the Galactic Center or the Galactic Plane. The same for the neutrino sample studied in \cite{arxiv1410}. 

In the data presented in Table \ref{tab:ICud}, an excess of downgoing (Down) events with respect to expectation seems however be present.
The Northern sky, inducing upgoing (Up) events in IceCube, contains only a small fraction of the Galaxy. 
For this reason, let us assume that the 3.6 estimated HESE from cosmic neutrinos ($n_{IC}$) arising from the Northern sky are all of extragalactic origin.
Assuming a $E^{-2.0}$ spectrum, a symmetric contribution from the North/South extragalactic sources and taking into account the different $A_{IC}^{\nu_i}$ for events coming from the North and South hemispheres, then 6.2 events are expected from the South (they reduce to 5.8 events for a $E^{-2.5}$ flux).
This excess of $\sim$7.5 events in the Southern sky corresponds to $\sim$ 40\% of the total signal.

\begin{table}[tb]
{\centering \begin{tabular}{l|c|c|c|c||l|c|c|c|c }
\hline 
HESE \cite{ic2} & Data & Bck & $n_{IC}$ & $N_{IC}$ & New  $\nu$ sample \cite{arxiv1410} & Data & Bck & $n_{IC}$ & $N_{IC}$\\
$E_{dep}>60$ TeV&      &   &  &   $E^{-2}$     & $E_{dep}>25$ TeV & & & & $E^{-2.46}$     \\ \hline \hline 
Up (North)	& 5	&1.4	& 3.6	&6.7	& Up $(\sin\delta>0.06$) & 11  & 5.3   &  5.7 & 12.1        \\
Down (South) &15	&1.3	&13.7	&11.5	&  Down $(\sin\delta<-0.06$) & 29  & 4.8   &   24.2 & 15.0       \\ \hline
All       	& 20	&2.7	&17.3	&18.2	&  All& 43  & 11.7    & 31.3 & 29.1         	\\ \hline 
\end{tabular}\par}
\caption{{\small Column 2 to 5: number of HESE with $ E_{dep}>$ 60 TeV (Data); number of background events (Bck); number $n_{IC}$  of signal events (Data-Bck); expected number $N_{IC}$  of signal events from the best-fit. 
Column 7 to 10: the same quantities for sample with $E_{dep}>25$ TeV in \cite{arxiv1410}.
Quantities are given for upgoing (from the Northern hemisphere) and downgoing (from the Southern hemisphere) events, and for the whole sky (All).
\label{tab:ICud}}}
\end{table}

An even stronger excess from the South is derived from data in \cite{arxiv1410}, using events with $E_{dep}>25$ TeV.
Here, to the 5.7 signal events coming from the North with $\sin\delta>0.06$ (assuming the same considerations of above) should correspond $\sim$ 7.2 events from the South with $\sin\delta<-0.06$.
As 24.2 events are observed, more than 50\% of the number of signal events in the whole sky seems to be produced by a non-isotropic cosmic component, likely of Galactic origin. 
A possible contribution from transient extragalactic objects located in the Southern sky can be considered as well.

The above conclusions are derived for $\Gamma=2.0$; however similar results are obtained using softer spectral indexes (i.e. $\Gamma>2.0$) for the cosmic neutrino flux. 
This is relevant because different models involving Galactic, extragalactic or exotic origin of the IC signal exist in the literature.
The neutrino flux predicted by each model has a preferred value, usually ranging in the interval $\Gamma= 2.0\div 2.7$. 

In the following, the effects of the hypothesis that a sizeable fraction of the cosmic neutrinos observed by IC is originated in our Galaxy is considered.
Following the methods reported in \cite{spu}, the neutrino flux from point-like or extended sources compatible with the above evaluated Galactic fraction of the IC signal is derived for different values of $\Gamma$. 

A signal originating from the Southern sky region can be observed by the ANTARES neutrino telescope \cite{anta}, located in the Mediterranean Sea.
Existing ANTARES upper limits derived under the hypothesis of a $\Gamma=2.0$ neutrino spectrum are used to infer upper limits for $\Gamma>2.0$. The ANTARES expected sensitivities for extended sources are used to discuss the conditions under which an IceCube hot spot can be observed.

\section{The ANTARES and IceCube effective areas}

The neutrino effective area at a given energy, $A_{eff}(E)$, is defined as the ratio between the neutrino event rate in a detector (units: s$^{-1}$) and the neutrino flux (units: cm$^{-2}$ s$^{-1}$) at that energy. 
The effective area depends on the flavor and cross-section of neutrinos, on their absorption probability during the passage through the Earth, and on detector-dependent efficiencies.
Detector efficiencies are correlated to each particular analysis, referring to the criteria used to trigger and to reconstruct the events, and to the cuts applied to reduce the background. 
A fraction of the irreducible background due to atmospheric neutrinos contaminates in any case the signal, with a percentage depending on the strength of cuts used to define $A_{eff}(E)$.

\begin{figure}[tb]
\begin{center}
\includegraphics[width=11.5cm]{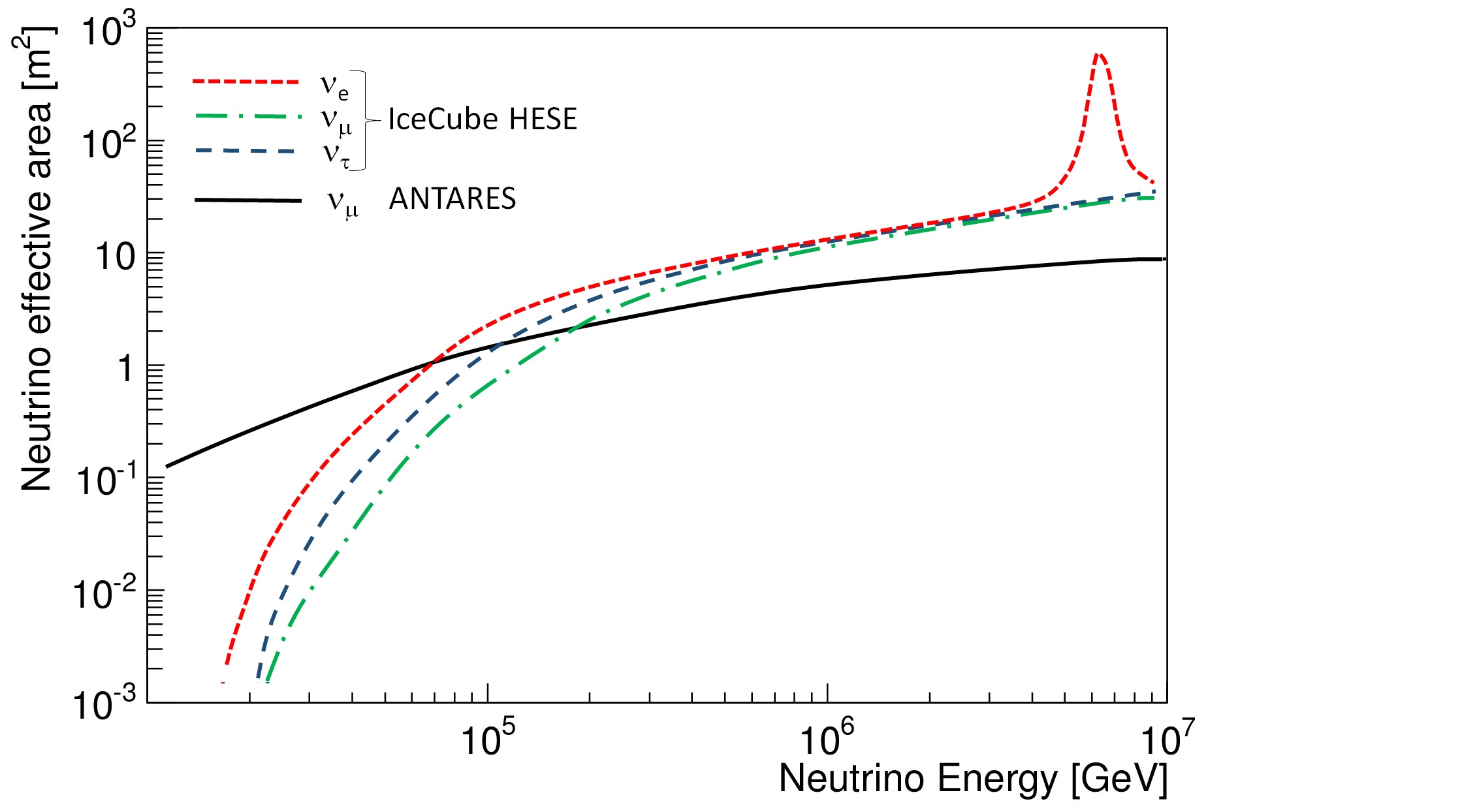}
\end{center}
\caption{\small\label{fig:Aeff} Full black line: ANTARES $\nu_\mu$ effective area \cite{anta12}. The neutrino track is determined with a median angle of $\lesssim 0.4^\circ$.
Colored dashed lines: IceCube $\nu_e$, $\nu_\mu$ and $\nu_\tau$ effective areas \cite{ic1} from the analysis yielding the HESE. The backgrounds due to atmospheric muons and neutrinos is largely suppressed above few tens of TeV.}
\end{figure}

Figure \ref{fig:Aeff} shows the ANTARES effective area ($A_{ANT}^{\nu_\mu}$, full black line) for the $\nu_\mu$ flavor as derived in the analysis \cite{anta12} for the search for cosmic neutrino point sources in the declination band containing the Galactic Center. 
The red ($A_{IC}^{\nu_e}$), green ($A_{IC}^{\nu_\mu}$) and blue ($A_{IC}^{\nu_\tau}$) lines refer to the IceCube analysis on a $4\pi$ sr yielding the HESE. 

Despite the fact that the ANTARES instrumented volume is much smaller than 1 km$^3$, Fig. \ref{fig:Aeff} shows that the ANTARES $\nu_\mu$ effective area is larger than $A_{IC}^{\nu_e}$, $A_{IC}^{\nu_\mu}$ and $A_{IC}^{\nu_\tau}$ below $\sim$ 60 TeV. 
At the highest energies where neutrinos were detected, 2 PeV, $A_{IC}^{\nu_\mu}$ is a factor of two larger than $A_{ANT}^{\nu_\mu}$ while the total IC effective area for HESE ($A_{IC}^{\nu_e} + A_{IC}^{\nu_\mu} +A_{IC}^{\nu_\tau}$) is only 7.3 times larger  than $A_{ANT}^{\nu_\mu}$.

These values of the $A_{eff}$ of the two experiments are largely dominated by the different criteria hidden in the analysis.
Strong cuts are used in the IC analysis to select a high-purity sample of diffuse high-energy cosmic neutrinos with interaction vertex inside the instrumented volume.
The estimated angular resolution is $\sim 1^\circ$ for $\nu_\mu$ and $\sim 10^\circ -15^\circ$ for neutrino interactions producing showers (mainly from charged current interactions of $\nu_e,\nu_\tau$).
The criteria used in the ANTARES analysis allow a larger contamination of lower-energy atmospheric neutrinos, but enable to reconstruct $\nu_\mu$ events with superior angular resolution, $\sim 0.4^\circ$.
The consequence is that ANTARES has equivalent (or superior, depending on the signal spectral index) capability to extract a signal if the cosmic source is located in the Southern sky, and it is point-like or confined in a region seen within a small solid angle $\Delta\Omega$ by the detector.

\section{Normalization factors for different cosmic spectral indexes}
The standard diffusive shock acceleration model yields a $\Gamma=2.0$ spectral index for primary CRs, and consequently for secondary $\gamma$-rays and neutrinos. 
However, most $\gamma$-ray sources observed in the GeV and TeV range show spectral indexes larger than 2.0.  
The reason why $\gamma$-ray spectra from supernovae remnants are observed with spectral indexes $\Gamma\simeq 2.2-2.3$ remains unclear. 
A softer spectral index ($\Gamma \simeq 2.4-2.5$) is consistent with the theoretical model of CR injection by diffusive shock acceleration followed by escape through the Galactic magnetic field with Kolmogorov turbulence \cite{nese}.

Thus, it is important to consider the normalization factors $\Phi^{D,\Gamma}_0$ for the IC signal for different models of cosmic fluxes ${E^\Gamma } \Phi^{D,\Gamma} (E) \equiv \Phi_0^{D,\Gamma} (E)$ (in units: GeV cm$^{-2}$ s$^{-1}$ sr$^{-1}$. The $D$ stands for $diffuse$.)
The same number $N_{IC}$ of events for different $\Phi^{D,\Gamma} (E) $ is obtained using the effective area  $A_{IC}(E)\equiv [A_{IC}^{\nu_e}+ A_{IC}^{\nu_\mu}+ A_{IC}^{\nu_\tau}]$ and detector livetime $T$:
\begin{equation}\label{eq:ic2}
N_{IC} = T \cdot \int \Phi^{D,\Gamma} (E) \cdot A_{IC}(E) \cdot dE\cdot d\Omega 
    = 4\pi T \cdot \Phi^{D,\Gamma}_0 \cdot \int   E^{-\Gamma} \cdot A_{IC}(E)\cdot dE 
= 4\pi T \cdot \Phi^{D,\Gamma}_0 \cdot {\cal D}_\Gamma \ .
\end{equation}
The integral ${\cal D}_\Gamma$ (the \textit{detector response})  extends over the energy range where $ A_{IC}(E)$ is not null, and is computed numerically. 

Let us assume that $n_p$ events out of $N_{IC}$ are produced by a point-like source with generic spectrum: 
$E^\Gamma \Phi^{p,\Gamma} (E) = \Phi^{p,\Gamma}_0 \quad \textrm{ (units: GeV cm}^{-2} \textrm{ s}^{-1} \ .$ The $p$ stands for $point-like$.)
The normalization factor $\Phi^{p,\Gamma}_0$ necessary to produce $n_p$ events is obtained by requiring that:
\begin{equation}\label{eq:gal3}
n_p = T \cdot \int \Phi^{p,\Gamma} (E) \cdot A_{IC}(E)\cdot dE 
    = T \cdot \Phi^{p,\Gamma}_0  \int E^{-\Gamma} \cdot A_{IC}(E)\cdot dE 
    = T \cdot \Phi^{p,\Gamma}_0 \cdot {\cal D}_\Gamma
\end{equation}
where $T$, $ A_{IC}(E)$ and, consequently, the detector response ${\cal D}_\Gamma$ are the same as in Eq. (\ref{eq:ic2}). 
Then, the normalization factor for a point-like source flux of a given spectral index $\Gamma$ is given by:
\begin{equation}\label{eq:gal5}
\Phi^{p,\Gamma}_0 = 4\pi\cdot \biggl({n_p \over N_{IC}}\biggr) \cdot \Phi_0^{D,\Gamma} \ .
\end{equation}

If a fraction $ n_{\Delta \Omega} $ of the IceCube signal is produced in a region of the Southern sky of angular extension $\Delta \Omega\ll 4\pi$ sr, and flux $E^{\Gamma} \Phi^{D^\prime,\Gamma} (E) = \Phi^{D^\prime,\Gamma}_0$, the  signal can be observed as an \textit{enhanced diffuse flux}.
Similarly to Eq. (\ref{eq:gal5}), using the detector response derived in (\ref{eq:ic2}), we obtain 
\begin{equation}\label{eq:gad3}
\Phi^{D^\prime,\Gamma}_0 = \biggl({n_{\Delta \Omega} \over N_{IC}}\biggr) \cdot \biggl({4\pi \over {\Delta \Omega}}\biggr) \cdot \Phi_0^{D,\Gamma} \ .
\end{equation}

\begin{table}[tb]
{\centering \begin{tabular}{c||c|c|c|c|c||c}
\hline 
&  \multicolumn{6}{c}{{ units: (GeV cm$^{-2}$ s$^{-1}$)  }} \\

& \multicolumn{5}{|c||}{{ $\Phi^{p,\Gamma}_0$ (from HESE) }} & ANTARES \\
$\Gamma=$ & $n_p=1$ &$n_p=2$ &$n_p=3$ &$n_p=4$ &$n_p=5$ & 90\% C.L. limit
 \\ \hline
2.0 & $6.9\  10^{-9}$ & $1.4\  10^{-8}$ & $2.1\  10^{-8}$ & $2.8\  10^{-8}$ &  {$3.5\  10^{-8}$} & $4.0\  10^{-8}$ \\
2.2 & $9.0\  10^{-8}$ & $1.8\  10^{-7}$ & $2.7\  10^{-7}$ & \underline{$3.6\  10^{-7}$} & - & $3.2\  10^{-7}$ \\
2.3 & $3.3\  10^{-7}$ & $6.6\  10^{-7}$ & \underline{$9.9\  10^{-7}$} & - & - & $8.4\  10^{-7}$ \\
2.4 & $1.2\  10^{-6}$ & \underline{$2.3\  10^{-6}$} & - & - & - & $2.2\  10^{-6}$ \\
\hline
\end{tabular}\par}
\caption{{\small Column 2 to 6: normalization factors $\Phi^{p,\Gamma}_0$ yielding  $n_p=1,.. .,5$ HESE in IceCube vs. $\Gamma$. The last column shows the 90\% C.L. upper limits for a $\Gamma=2.0$ point-like source derived from ANTARES \cite{anta14}. The values for $\Gamma>2.0$ were derived in \cite{spu}. The first value in each row excluded by these limits is underlined.
\label{tab:point}}}
\end{table}

\section{ANTARES constraints for the IC signal from the Southern sky}\label{sez:anta}

\textbf{Point-like sources.}
Table \ref{tab:point} reports the normalization factor $\Phi^{p,\Gamma}_0$ for a point-like source necessary to produce $n_p=1\div 5$ HESE, as derived from Eq. (\ref{eq:gal5}). Four different values of $\Gamma$ are considered.
Point-like sources in the Galactic central region were searched for by ANTARES \cite{anta14} and upper limits as a function of the source declination were derived assuming a spectral index $\Gamma=2.0$.
Following the procedure defined in \cite{spu}, the the 90\% C.L. upper limit for a point-like source has been translated to upper limits for softer spectral indexes. The ANTARES results for $\Gamma=2.0$ in the Galactic Center region and the derived valued for $\Gamma=$2.2, 2.3 and 2.4 are reported in the last column of Table \ref{tab:point}.
The ANTARES 90\% C.L. upper limit excludes a single point-like source with $\Gamma=2.0$ producing more than 5 HESE. 
The derived limit excludes a single point-like source yielding a cluster of more than 2 events for $\Gamma=2.3$, while the presence of a cluster made of two or more events is excluded for $\Gamma>2.3$.

\vskip 0.2cm
\noindent \textbf{Enhanced diffuse flux.}
Table \ref{tab:diffu} (columns from 3 to 6) shows the normalization factors from Eq. (\ref{eq:gad3}), assuming $ n_{\Delta \Omega}= 3 \div 6$ HESE within a solid angle region $\Delta \Omega=2\pi(1-\cos\theta)$ corresponding to a circular windows of  $\theta=8^\circ$.
The ANTARES strategy for the study of an enhanced diffuse flux is different with respect to that for the search for point-like sources.
This latter relies mainly on the pointing accuracy of the telescope.
The expected background due to atmospheric neutrinos within a circular windows of $\theta\lesssim 1^\circ$ is small and this is not anymore true for larger values of $\theta$.
As the energy spectrum from a cosmic signal (either point-like or diffuse) is expected to be harder than that of atmospheric neutrinos, the signal should exceed the background above a certain threshold of the reconstructed energy.
Thus, the discrimination between signal and background needs the use of the estimated energy of the event, similarly to the case of the search for a diffuse flux of high energy $\nu_\mu$ \cite{anta_diffu}.

ANTARES has used an Artificial Neural Network to estimate the energy of the muons entering the detector for studying the Fermi bubbles (FB) \cite{anta-fb}.
The reported ANTARES sensitivity in terms of an enhanced diffuse flux from the FB region, assuming a $E^2 \Phi(E)$ spectrum without cutoff up to the PeV energies is $3.1\times 10^{-7}$ GeV cm$^{-2}$ s$^{-1}$ sr$^{-1}$. 
In the analysis, using 806 days livetime, 16 events were found, with an expected background of 11 events. For $\Gamma=2.0$ the background corresponds to 7.5 events/(sr $\cdot$ y). 
The derived 90\% C.L. upper limit is $E^2 \Phi^{FB}(E)=5.4\times 10^{-7}$ GeV cm$^{-2}$ s$^{-1}$ sr$^{-1}$.
As the sensitivity depends on the background rate, different optimizations must be deduced for different spectral indexes; in general, it can be assumed that the background level slightly increases for softer spectral indexes \cite{spu}.
The sensitivities extrapolated from the ANTARES FB analysis for $\Gamma>2.0$ are reported in the last column of Table \ref{tab:diffu}. 
According to these values, a dedicated search for a directional neutrino flux, for instance around the IC hot spot, would produce a positive result for any spectral indexes $\Gamma \ge 2.0$, if $\Delta \Omega\le 0.06$ sr (or circular window of $\theta<8^\circ$) and $n_{\Delta \Omega}>2$. 
For a signal spread out on a larger circular window, the minimum sensitivity would correspond to a higher $n_{\Delta \Omega}$.

\begin{table*}[tb]
{\centering \begin{tabular}{c|c||c|c|c|c||c}
\hline 
& & \multicolumn{5}{c}{{ units: (GeV cm$^{-2}$ s$^{-1}$ sr$^{-1}$)  }} \\
$\Delta \Omega$ & & \multicolumn{4}{|c||}{{ $\Phi^{D^\prime,\Gamma}_0$ (from HESE) }} & ANTARES \\
(sr) & $\Gamma=$ & $ n_{\Delta \Omega}=3$ & $ n_{\Delta \Omega}=4$ &$ n_{\Delta \Omega}=5$ &$ n_{\Delta \Omega}=6$  & sensitivity  
 \\  \hline
0.06 & 2.0 & $3.5\ 10^{-7}$ & $4.6\ 10^{-7}$ & $5.8\ 10^{-7}$ & $7.0\ 10^{-7}$ & $3.1\ 10^{-7}$ \\
  & 2.2 & $4.5\ 10^{-6}$ & $6.0\ 10^{-6}$ & $7.5\ 10^{-6}$ & $9.0\ 10^{-6}$  & $3.6\ 10^{-6}$ \\
  & 2.3 & $1.7\ 10^{-5}$ & $2.2\ 10^{-5}$ & $2.8\ 10^{-5}$  & $3.3\ 10^{-5}$  & $1.1\ 10^{-5}$ \\
  & 2.4 & $5.9\ 10^{-5}$ & $7.8\ 10^{-5}$ & $9.8\ 10^{-5}$  & $1.2\ 10^{-4}$  & $3.4\ 10^{-5}$ \\
\hline
\end{tabular}\par}
\caption{{\small Column 3 to 6: Normalization factors $\Phi^{D^\prime,\Gamma}_0$ for an enhanced diffuse flux, obtained assuming $n_{\Delta \Omega}=3$ to 6 HESE in a circular window of 8$^\circ$ ($\Delta\Omega=0.06$ sr). In the last column, the value for $\Gamma=2.0$ corresponds to the ANTARES sensitivities from the FB regions \cite{anta-fb}. The sensitivities for $\Gamma>2.0$ are obtained in \cite{spu}. 
\label{tab:diffu}}}
\end{table*}

\vskip 0.2cm
\noindent \textbf{Regions of large angular size (Fermi bubbles, Galactic plane).}
Recent predictions of the neutrino flux from the FB regions \cite{luna} allow to estimate the expected number of events for the IC detector assuming the effective area of the HESE.
By folding the predicted $\nu$ spectra with the ANTARES effective area, the number of $\nu_\mu$ induced events in ANTARES would correspond to $\sim 30\%, 50\% $ and 100\% of the $\nu_e+\nu_\mu+\nu_\tau$ HESE in the same livetime, for $\Gamma=2.0, 2.1$ and $2.3$, respectively. However, this corresponds to a smaller ANTARES sensitivity with respect to IC, due to the larger background induced by atmospheric neutrinos from the wide FB region ($\Delta \Omega\sim 0.8$ sr).

The excess of HESE events from the Galactic region could finally be produced by interaction during propagation of freshly injected CRs with spectral index $\Gamma \simeq 2.4-2.5$ \cite{nese}.
The preliminary ANTARES upper limits from the Galactic plane are reported in \cite{visser}.

\section{Conclusions}\label{sez:disc}

The ANTARES detector has sufficient sensitivity to test many models that explain a fraction of the HESE sample in IceCube in terms of a Galactic component.
Models in which more than 2 HESE  are originated from a point-like and steady source in the Southern hemisphere are excluded for spectral indexes $\Gamma\ge 2.3$.
The possibility that a clustering of events is produced in a region of small angular size ($\Delta \Omega\simeq 0.1-0.2$ sr) in (or near) the Galactic Plane is under investigation in ANTARES.
As reported in Table \ref{tab:diffu}, the estimated ANTARES sensitivity is below the signal level (allowing a positive detection) for any spectral indexes $\Gamma \ge 2.0$, if $\Delta \Omega\le 0.06$ sr (i.e. a circular window of $\theta<8^\circ$) and $n_{\Delta \Omega}>2$. 
For a signal spread out on a larger solod angle, the minimum sensitivity would correspond to a higher $n_{\Delta \Omega}$.
For very large regions (the FB, the Galactic plane) the present sensitivities using the $\nu_\mu$ channel alone are above the model predictions.
The inclusion of showering events, with a relatively looser angular precision, would significantly increase the ANTARES sensitivities for the study of extended regions.

\vskip 0.2 cm
\noindent\textbf{\large Acknowledgments}

\noindent I would like to thank many members of the ANTARES and KM3NeT Collaborations for comments and in particular J. Brunner and A. Kouchner.


\end{document}